\documentclass{article}

\usepackage[preprint]{neurips_2026}

\usepackage[utf8]{inputenc} 
\usepackage[T1]{fontenc}    
\usepackage{hyperref}       
\usepackage{url}            
\usepackage{booktabs}       
\usepackage{amsfonts}       
\usepackage{nicefrac}       
\usepackage{microtype}      
\usepackage{xcolor}         
\usepackage{amsmath}
\usepackage{graphicx}
\usepackage{multirow}
\usepackage{tcolorbox}
\usepackage{listings}
\usepackage{subcaption}
\tcbuselibrary{skins,breakable}
\usepackage{float}

\newtcolorbox{promptcodebox}[1]{
  colback=gray!5, colframe=gray!50!black,
  fonttitle=\bfseries, title={#1},
  breakable, enhanced, boxrule=0.4pt, arc=1mm,
  left=4pt, right=4pt, top=4pt, bottom=4pt
}

\newtcolorbox{boxbase}[1]{
  colback=gray!5, colframe=blue!40!black,
  fonttitle=\bfseries, title={#1},
  breakable, enhanced, boxrule=0.4pt, arc=1mm,
  left=4pt, right=4pt, top=4pt, bottom=4pt
}

\newtcolorbox{boxpositive}[1]{
  colback=gray!5, colframe=green!50!black,
  fonttitle=\bfseries, title={#1},
  breakable, enhanced, boxrule=0.4pt, arc=1mm,
  left=4pt, right=4pt, top=4pt, bottom=4pt
}

\newtcolorbox{boxnegative}[1]{
  colback=gray!5, colframe=red!50!black,
  fonttitle=\bfseries, title={#1},
  breakable, enhanced, boxrule=0.4pt, arc=1mm,
  left=4pt, right=4pt, top=4pt, bottom=4pt
}
\makeatother

\newcommand{\code}[1]{\texttt{#1}}

\title{If It's Not Buggy, Don't Fix It: On the Dynamics of Iterative Bug-fixing with LLMs}

\author{%
  Xietao Wang-Lin\thanks{Work done during an internship at UnlikelyAI.} \\
  University of Warwick \\
  \And
  Anton Isopoussu \\
  UnlikelyAI \\ 
  \And 
  Louis Mahon \\
  UnlikelyAI
}

\begin{document}

\maketitle


\begin{abstract}
    Large language models (LLMs) have become ubiquitous in software development, with LLM-based automated program repair tools increasingly used during code review. In this report, we explore the iterative blind use of LLMs as bug-fixers. Across multiple models and repair environments, we find that LLMs consistently claim to detect bugs in entirely bug-free programs while the rate of repair of buggy programs is less than that of the damage to correct programs. We also explore the long-term dynamics of this iterative process, and find that this frequently reaches a pseudo-bug-fixing cycle where the same changes are added and removed again ad infinitum. Lastly, via mechanistic probing, we unveil the existence of a steering vector which controls the editing propensity, suggesting that LLMs have an internal representation of ``buggy code", and that this representation is what is falsely activated to induce pseudo-bug fixing. These results provide insight towards the dynamics of fully autonomous bug-fixing systems, as well as stopping conditions under ambiguous goals.
\end{abstract}
\section{Introduction}

Large language models (LLMs) are now ubiquitous in coding tasks. Increasingly, they are used as fully autonomous agentic harnesses, where the user gives little to no supervision and the agent iteratively processes the code, with the stopping condition being determined by the harness \citep{sweagent}. A similar loop arises in multi-agent interactions, where the output generated turns into the input for the next generation \citep{du2024improving, ko2026attractorstatesemergemultiturn}. However, little is known about the long-term effects of this iterative process, such as code correctness or how and when the agent decides to terminate the iteration, especially when there is no objective measure of task performance.

Bug-finding is an essential part of the coding process. Neither human or LLM code are rarely perfect on the first attempt, and detecting and fixing potential errors is crucial. LLMs often do this via automated bug-fixing agents that operate iteratively, i.e., the LLM searches for bugs, applies patches, then again searches the patched code and potentially applies more fixes, iterating until no more bugs are found \citep{liu2024marscodeagentainativeautomated, repairagent}. 
Given the ubiquity and importance of this setup, some natural questions arise: how does code evolve under iterative bug-fixing? Are LLMs aware of bugs under vague conditions? When is the LLM satisfied the code is completely bug-free?

In this report, we particularly focus on iterative bug-fixing using LLMs, where LLMs review and fix bugs on the same piece of code until it finds no more bugs. We consider the case where the prompt does not include history, that is, the agent sees the current state of the code, but not the sequence of states the code has been in through past bug fixes \citep{agentless}. This situation may arise, for instance, when a user approves changes blindly, when including past states would exceed the context window or in multi-agent setups with delegation \citep{usersinsecure, lostllms}.

Our contributions are as follows:
\begin{itemize}
    \item We study iterative bug-fixing dynamics and quantify the rate at which it improves buggy code, as well as the rate at which it damages correct code by trying to remove pseudo-bugs;
    \item We show that, under relatively standard settings, the damage rate can end up being significantly higher than the repair rate;
    \item We document the behaviour of pseudo-bug-fixing cycles, which appear more frequently when using atomic changes compared to whole-file edits.
    \item By mechanistic probing, we unveil the existence of a ``buggy code" internal representation in LLMs, which is activated when the LLM detects bugs (either pseudo-bugs or real bugs) and not when the LLM thinks the code is bug-free.
\end{itemize}

\section{Related work}
\textbf{Dynamical systems in LLMs}. Recent work often takes a dynamical system view to model a variety of behaviours of LLMs. \citet{shumailov_ai_2024} consider recursive training as a Markov chain over the empirical data distribution. Looped transformers can be seen as a dynamical system over the looped latent states, and have been shown to have cycles and fixed points \citep{geiping2025scaling, movahedi2026fixedpointreasonersstableadaptive, blayney2026mechanisticanalysisloopedreasoning}.
The framework of dynamical systems can be used to study iterative generation, where the output is recirculated into the input of the next LLM, creating a dynamical system over the text generations. This has  previously been studied in the literature in the form of iterative paraphrasing \citep{geng2026markoviangenerationchainslarge, wang-etal-2025-unveiling}, backtranslation \citep{geng2026markoviangenerationchainslarge, mohamed-etal-2025-llm, hong-etal-2025-consistencychecker} and program mutation \citep{gurkan26mutation}, where LLMs converge to attractors or periodic states. \citet{perez2025when} look into text semantics and find that iterative generation over texts has specific cultural attractors. \citet{wu2026languagemodelsconvergethemselves} study iterative refinement of paper abstracts and find that this process converges exponentially fast.

\textbf{Iterative coding with LLMs}. Similar iterative generation loops have been studied in the context of coding tasks. \citet{choi2026anchoring} used a bug-generator together with a bug-fixer LLM to improve bug-fixing via RL; \citet{peitek2026restructuringstabilizationlargescaleexperiment} repeatedly prompt an LLM to improve code readability, converging to a fixed-point in a few iterations, thereby suggesting that LLMs might have an ``optimal readability” internal representation of code. Self-refine has also been applied to code snippets \citep{madaan2023self, olausson2024is}, where the self-refinement loop is used to improve the pass rate of code solutions. \citet{gao2026loopingreliabilitystateboundevidence} deviate from the self-refine loop by giving the LLM evidence from the code rather than feedback, finding that regression bugs are common in generation-test-revise loops.

\textbf{Long-horizon tasks}. \citet{laban2026llmscorruptdocumentsdelegate}, using a chain of reversible tasks as toy model, found that most LLMs undertaking long-horizon tasks often make a critical mistake which they never recover from. Similarly, \citet{orlanski2026slopcodebench}, constructed a benchmark where LLMs iteratively extend their previous code, and find that correctness and conciseness degrade over time.

\textbf{Steering activations}. 
The linear representation hypothesis \citep{park2024the} poses that high-level concepts are represented linearly, and thus, linear probes have become a popular method to capture the underlying concepts \citep{marks2024the}. The vectors obtained from the probe, known as steering vectors, can be used to steer the model to amplify or lessen the captured concept during generation \citep{wu2025axbench}. 
Indeed, recent work suggests that LLMs linearly encode concepts such as confidence \citep{venhoff2025understanding} and the overall likelihood of success of their current trajectory \citep{xu2026sparserewardsubsystemlarge, jiang2026valueaxislanguagemodels}. In this report, we use linear probes to obtain a steering vector of ``code bugginess", which controls the iterative bug-fixing loop.

\section{Iterative bug-fixing}
\begin{figure}
    \centering
    \includegraphics[width=\linewidth]{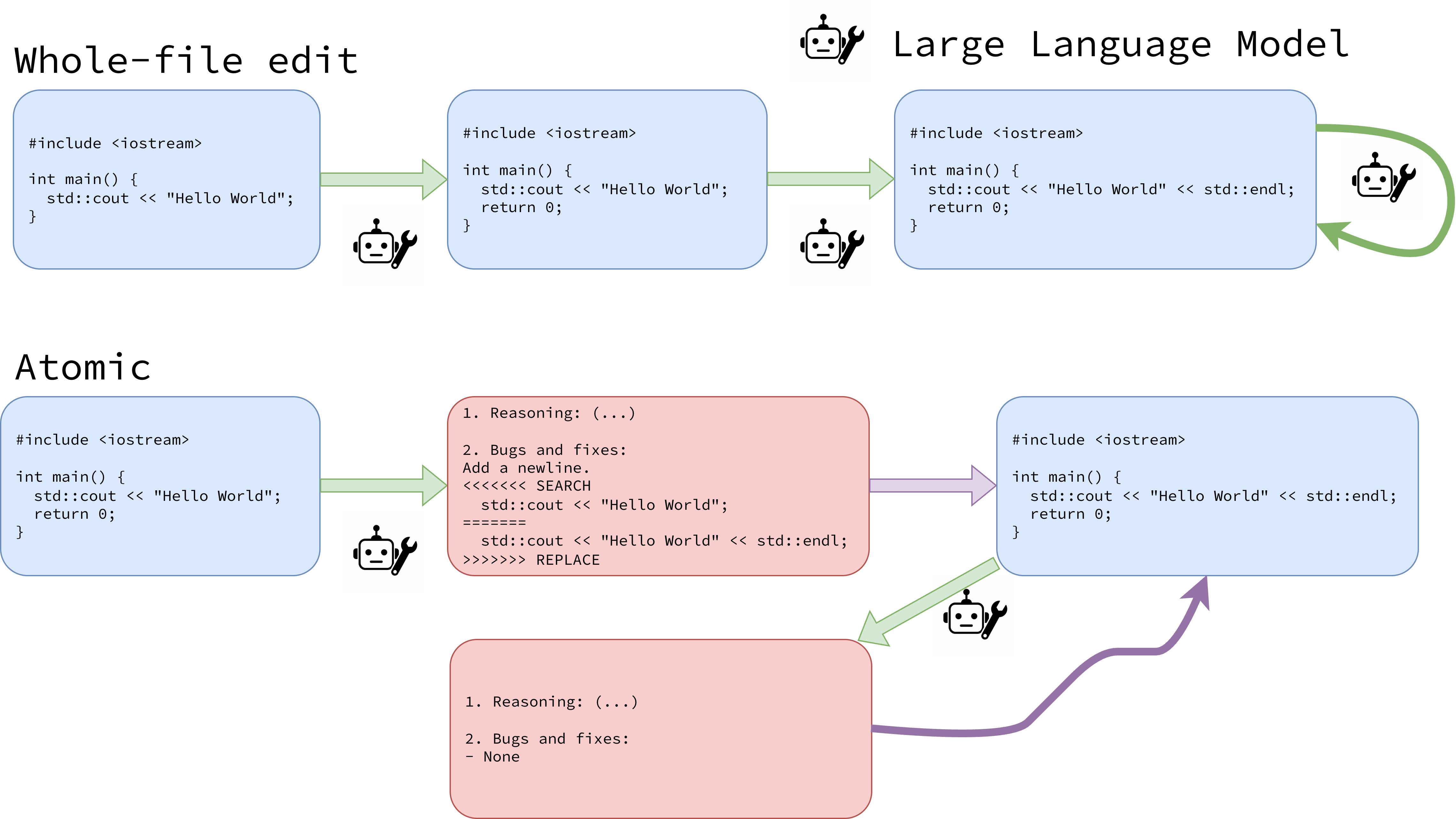}
    \caption{Diagram of the iterative bug-fixing loop. (Top) Using whole-file edits, the LLM takes in the input code file and outputs the corrected code, which is then used as the next input. (Bottom) Loop using atomic changes, the LLM reads the code file and outputs a reasoning trace, followed by one search/replace block, which is applied programmatically. The resulting code after the atomic change application is then used as input for the turn.}
    \label{fig:diagram}
\end{figure}

\textbf{Prompts}. We evaluate two techniques of code repair: (1) whole-file editing, where the LLM output is the entire repaired code, and (2) search/replace blocks (SRBs), as our style for atomic changes. SRBs, which have been used in Aider \citep{aider}, SWE-RL \citep{wei2025swerl} and Agentless \citep{agentless}, are preferred over formats involving line numbers (e.g. unified-diffs) in this setup as we only have a single code file to edit and it has been shown in previous literature that LLMs often make mistakes on line numbers, often requiring an additional post-processing step to align intended lines \citep{sun2026debugharnessemulatinghumandynamic}. In this case, we prompt the LLM to only fix the single most important bug (i.e. one bug at a time); as small code files often have overlapping logic, and we found that the LLM often suggests overlapping SRBs.
Exact prompts can be found in Appendix \ref{app:prompts}.

\textbf{Bug-fixing as a dynamical system}. Iterative generation can be viewed as a discrete dynamical system. Consider the space of all possible code files $\mathcal{C}$. If using greedy decoding, an instance of bug-fixing is represented by a fixer function, $f\colon \mathcal{C}\to \mathcal{C}$, such that $f$ is an instruction-tuned LLM and a prompt. A diagram of the setup is provided in Figure \ref{fig:diagram}.

Then, given an initial seed code, $C_0 \in \mathcal{C}$, we can iteratively fix the code using the recurrence,

\begin{equation}
    C_{n+1} = f(C_n), \quad n \in \{0, 1, 2, ..., N\}.
\end{equation}

When using SRBs, the code update comes from programmatically applying the SRB produced by the model. If the block is invalid (wrong format, search block not in code), we apply an identity change.
Hence, the fixer function is now, $f\colon \mathcal{C} \to \mathcal{B}$, where $\mathcal{B}$ is the space of all possible search/replace blocks, and we define $g\colon \mathcal{C} \times \mathcal{B} \to \mathcal{C}$, as the function that applies the SRB. For $B \in \mathcal{B}$, let

\[
g(C,B)
=
\begin{cases}
\operatorname{apply}(C,B), & \text{if } B \text{ is valid for } C,\\[4pt]
C, & \text{if } B \text{ is invalid for } C.
\end{cases}
\]

Next, we define \[
\hat{g}(C) = g(C, B),\quad B = f(C),
\]
and
\[
C_{n+1} = \hat{g}(C_n) =
g(C_n,B_n), \quad B_n = f(C_n), \quad n \in \{0, 1, 2, ..., N\}.
\]

Using stochastic decoding, such as top-$k$ and top-$p$, $f$ is now a Markov transition kernel, that is,

\begin{equation}
    \text{(whole-file) }C_{n+1} \sim  f(\cdot \mid C_n), \quad \text{or} \quad \text{(SRB) }B_{n} \sim f(\cdot \mid C_n),  \quad n \in \{0, 1, 2, ..., N\}.
\end{equation}

When using stochastic generation, for all our models, we set temperature $ \tau = 0.7$, top-$p$ to 0.9 and limit our generations to $N=100$ turns.

Note, given that we do not append history or previous code iterations in the prompt, all our dynamical systems are strictly Markovian. 

\section{Experimental Evaluation}
\label{sec:experiments}
\textbf{Dataset}. We use CodeContests+ \citep{wang2025codecontestshighqualitytestcase}, a dataset that contains user submissions of competitive programming problems with an extended set of test cases
 containing both correct and incorrect code submissions. 
We say that a piece of code is correct if it passes all the hidden test cases in the allocated time limit, and incorrect otherwise; we do not use partial credit.

\begin{table}[h]
\centering
\caption{Lines of code (LOC) statistics of the user submissions considered in this report.}
\label{tab:data}
\begin{tabular}{lccccccc}
\toprule
Submission type & Median LOC & Min. LOC & Max. LOC  \\
\midrule
Incorrect & 38 & 5   & 189 \\
Correct   &  42 & 12  & 119 \\
\bottomrule
\end{tabular}
\end{table}

Throughout the report we always use the same 20 problems, which are chosen at random; and from each problem, the same 40 C++ user submissions, also chosen at random. The average number of test cases is 23, and the average time-limit is $2.1$ seconds.

\textbf{Models}. We evaluate Gemini 2.5 Flash-Lite \citep{comanici2025gemini25pushingfrontier}, and Qwen2.5-7B-Instruct \citep{qwen2.5}.

\subsection{Repair Rate and Damage Rate}

First, we compute the empirical repair rate, $\alpha$ and damage rate, $\beta$. That is,
\begin{equation}
    \alpha = \mathbb{P}(C_{t+1} \text{ is correct} \mid C_{t} \text{ is incorrect}),
\end{equation}
\begin{equation}
    \beta = \mathbb{P}(C_{t+1} \text{ is incorrect} \mid C_{t} \text{ is correct}).
\end{equation}
The computed empirical rates are shown in Table \ref{tab:damage-repair}.

\begin{table}[htbp]
\centering
\caption{Repair rate and damage rate across different configurations and their respective standard error of the mean. Column name, `correct' vs `incorrect', refers to the starting state of the trajectory, and $\alpha$ and $\beta$ are then computed over all transitions in that trajectory.}
\label{tab:damage-repair}
\small
\begin{tabular}{lllrrrr}
\toprule
& & & \multicolumn{2}{c}{Gemini 2.5 Flash-Lite} & \multicolumn{2}{c}{Qwen2.5-7B-Instruct} \\
\cmidrule(lr){4-5} \cmidrule(lr){6-7}
& & & SRB & Whole-file & SRB & Whole-file \\
\midrule

\multirow{4}{*}{Correct}
& \multirow{2}{*}{$\tau=0$}   & $\alpha$ & $0.062 \pm 0.005$ & $0.040 \pm 0.009$ & $0.041 \pm 0.004$ & $0.016 \pm 0.009$ \\
&                          & $\beta$  & $0.293 \pm 0.011$ & $0.165 \pm 0.009$ & $0.424 \pm 0.015$ & $0.099 \pm 0.010$ \\
& \multirow{2}{*}{$\tau=0.7$} & $\alpha$ & $0.020 \pm 0.001$ & $0.003 \pm 0.000$ & $0.005 \pm 0.000$ & $0.008 \pm 0.001$ \\
&                          & $\beta$  & $0.216 \pm 0.004$ & $0.015 \pm 0.001$ & $0.190 \pm 0.005$ & $0.016 \pm 0.001$ \\

\midrule

\multirow{4}{*}{Incorrect}
& \multirow{2}{*}{$\tau=0$}   & $\alpha$ & $0.023 \pm 0.002$ & $0.101 \pm 0.007$ & $0.004 \pm 0.001$ & $0.019 \pm 0.004$ \\
&                          & $\beta$  & $0.261 \pm 0.032$ & $0.073 \pm 0.015$ & $0.440 \pm 0.099$ & $0.063 \pm 0.043$ \\
& \multirow{2}{*}{$\tau=0.7$} & $\alpha$ & $0.007 \pm 0.000$ & $0.004 \pm 0.000$ & $0.000 \pm 0.000$ & $0.001 \pm 0.000$ \\
&                          & $\beta$  & $0.187 \pm 0.007$ & $0.005 \pm 0.001$ & $0.391 \pm 0.051$ & $0.003 \pm 0.001$ \\

\bottomrule
\end{tabular}
\end{table}

\begin{figure}[!htbp]
    \centering
    \includegraphics[width=0.8\linewidth]{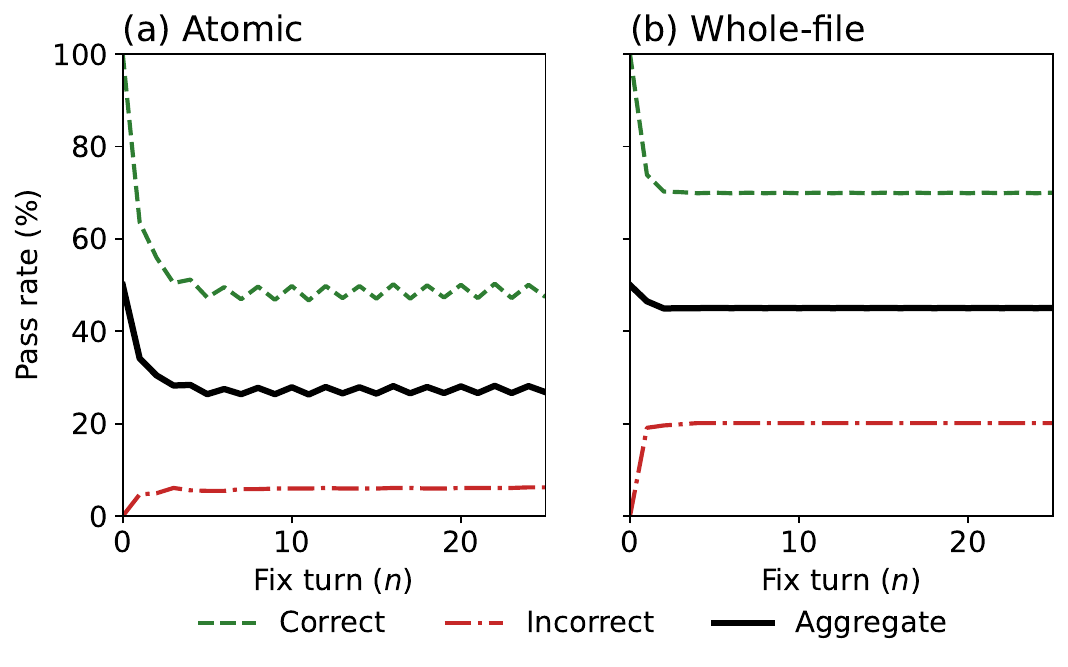}
    \caption{Average pass rate across turns using Gemini 2.5 Flash-Lite, when $\tau = 0$. Correct initial submissions in green, incorrect initial submissions in red, and the overall aggregate in black. (a) Using SRBs, (b) whole-file edits.}
    \label{fig:pass_rate}
\end{figure}

\begin{figure}[!htbp]
    \centering
    \includegraphics[width=0.8\linewidth]{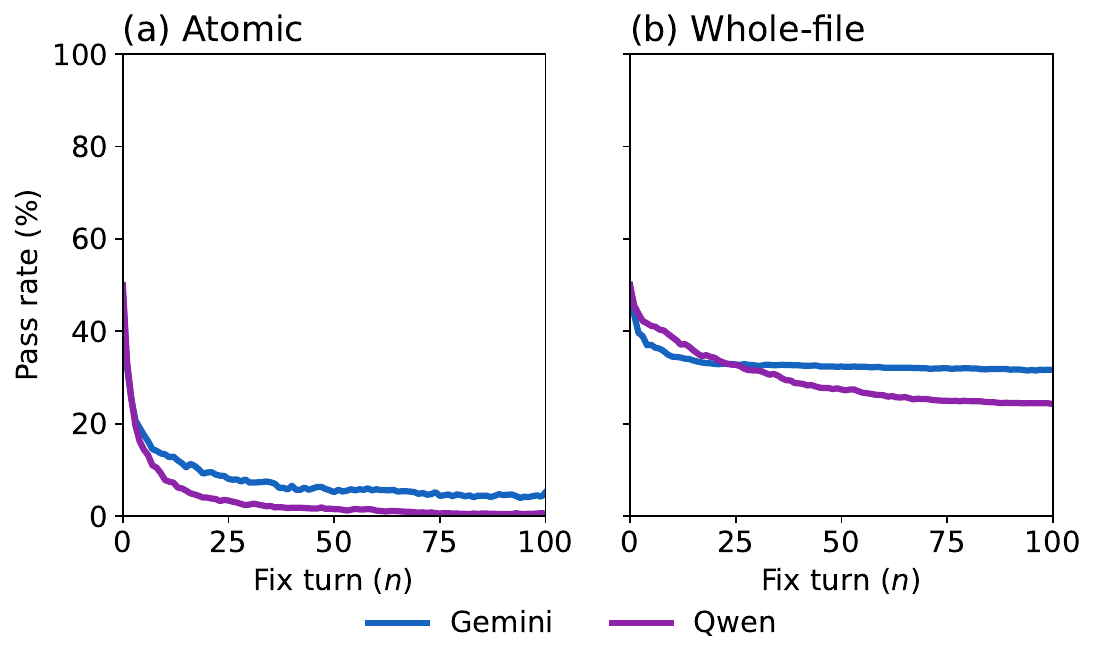}
    \caption{Average pass rate across turns when $\tau = 0.7$. Gemini 2.5 Flash-Lite in blue and Qwen2.5-7B-Instruct. (a) Using SRBs, (b) whole-file edits.}
    \label{fig:temperature}
\end{figure}

The same information is shown graphically, as trajectories, in Figures \ref{fig:pass_rate} and \ref{fig:temperature}. Overall, we see that the damage done by pseudo-bug fixing in correct files (damage rate) is at least as large as the improvements. For SRBs, the damage rate is substantially higher than the repair rate; for whole-file edits, the two are similar. The behaviour remains qualitatively similar in both $\tau=0$ and $\tau=0.7$.

\subsection{Attractor types}

\textbf{Attractor classification}. Under greedy decoding, if the fixer outputs a previously seen state in the chain, we stop the process as this leads to a loop. Loops of length-1 are fixed points, while longer loops are called cycles. States outside of attractors are denoted as transient, and the time to convergence to attractor as transient length.

When using SRBs, there are several ways a fixed point can be produced: 1) it outputs ``- None", meaning the fixer thinks there are no bugs, 2) an identity SRB, meaning the LLM says there is a bug to fix, but the fixed section is identical to the replaced section, 3) the search block cannot be located in the code, and 4) none of the above, usually this coincides with repetition degeneration \citep{Holtzman2020The} as comments, even though our prompt contains specific instruction not to add comments (see Appendix \ref{app:prompts}). On the other hand, when using whole-file edits, we only have two possible outcomes, either the output program has been seen previously during the bug-fixing loop or the output contains degenerate comments. Examples of possible attractor types for both SRBs and whole-file edits are shown in Appendix \ref{app:examples}. 

\begin{figure}[!htbp]
    \centering
    \includegraphics[width=0.8\linewidth]{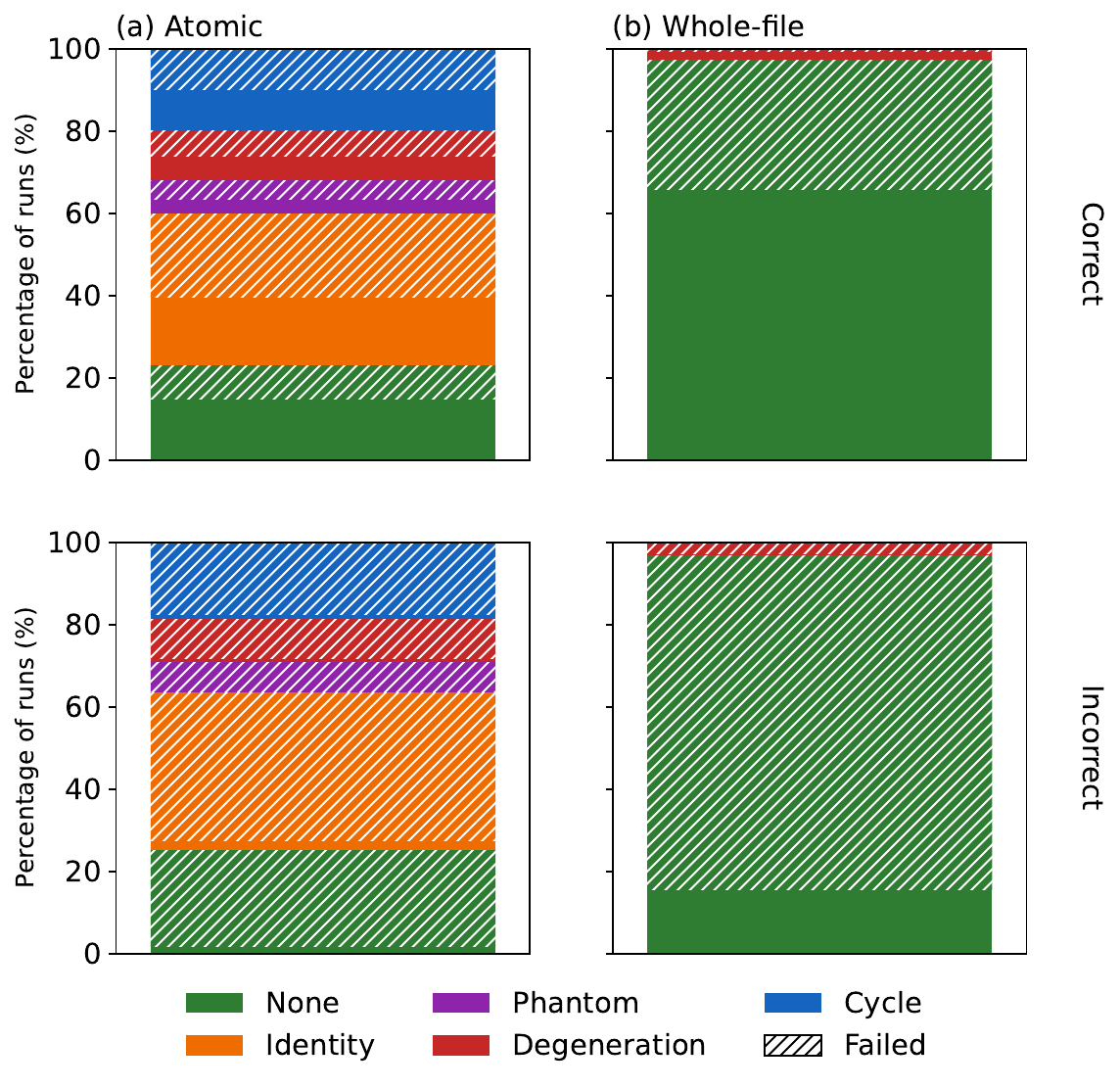}
    \caption{Attractor type frequency, using Gemini 2.5 Flash-Lite $\tau = 0$. (a) Using SRBs, (b) using whole-file edits. (Top) Starting with correct code. (Bottom) starting with incorrect code. We group whole-file edits identity output with None. Failed refers to submissions that are incorrect or do not compile.}
    \label{fig:attrator_types}
\end{figure}

\begin{figure}[!htbp]
    \centering
    \includegraphics[width=0.8\linewidth]{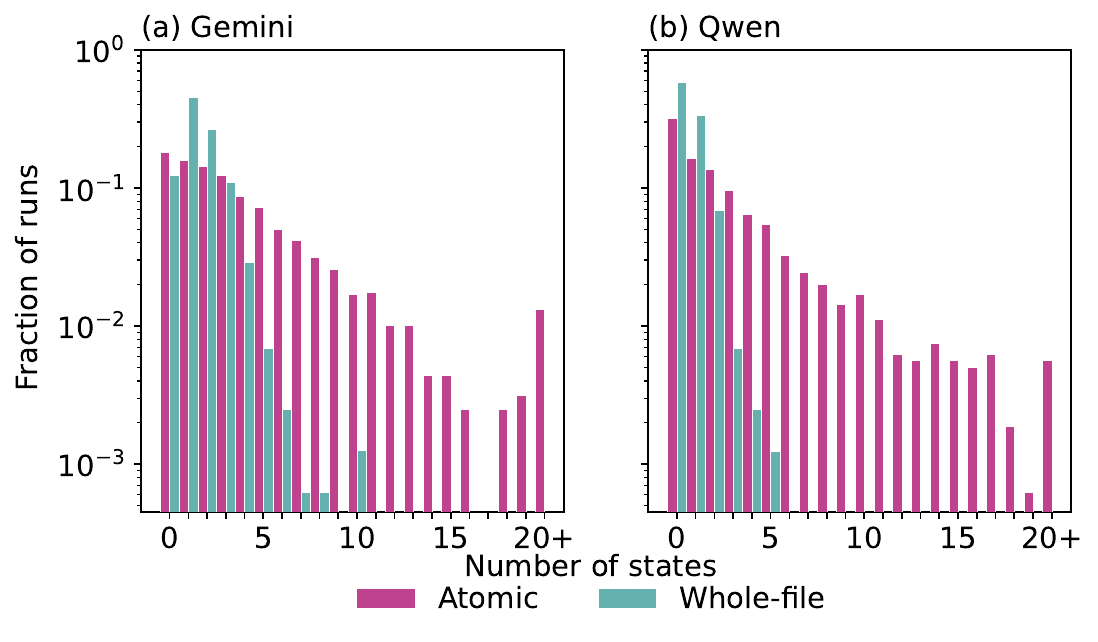}
    \caption{Transient length distributions when $\tau = 0$. (a) Gemini 2.5 Flash-Lite, and (b) Qwen2.5-7B-Instruct.}
    \label{fig:transient_dist}
\end{figure}

\begin{figure}[!htbp]
    \centering
    \includegraphics[width=\linewidth]{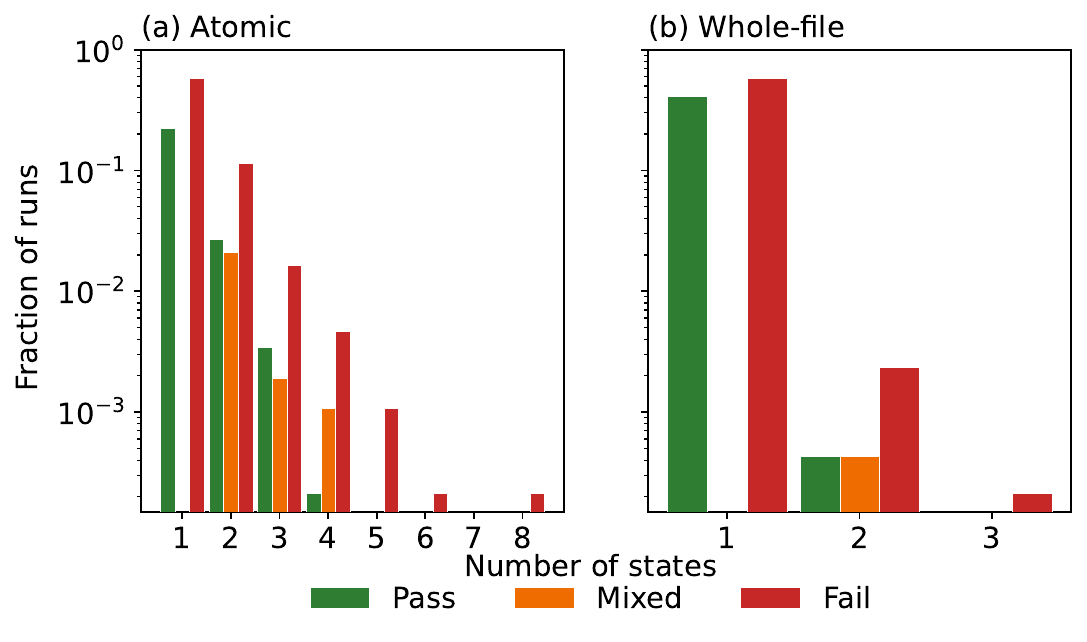}
    \caption{Cycle types for Gemini 2.5 Flash-Lite $\tau=0$. Pass (in green) are cycles with all correct states, mixed (in orange) are cycles with mixed correct/incorrect states, and fail (in red) are states with all incorrect states.}
    \label{fig:cycle_dist}
\end{figure}

Frequencies for each attractor type are shown in Figure \ref{fig:attrator_types}, transient length in Figure \ref{fig:transient_dist} and cycle length distributions are shown in Figure \ref{fig:cycle_dist}. The following observations arise: (1) SRBs induce significantly more cycles than whole-edits, (2) degenerations are more common when using SRBs as there are more opportunities for the model to degenerate, (3) SRBs produce significantly longer runs until convergence, likely due to the model only being allowed to fix one bug at a time, and (4) cycles are much longer when using SRB, also likely due to fixing one bug at a time. One possible explanation is that whole-file edits allow the model to maintain consistency of the overall logic, instead of having to make compensatory changes in a subsequent SRB. We also hypothesise that, of the many correct ways to write a given piece of code, the LLM often has an encoding of a single one. When it can rewrite the entire code, it simply selects this one correct way. If using SRBs instead, it tries to work off an alternative version of the code and can get confused even if that code is correct, because it does not meet the LLM's more narrow notion of correctness.

Lastly, as seen in Figure \ref{fig:cycle_dist}, cycles are not necessarily of one type (i.e. all the states are incorrect or all the states are correct). There are cycles with mixed correct/incorrect states in both SRBs and whole-file edits: for instance, when the cycle length is 2, this is a cycle where the loop is cycling between a correct and incorrect state infinitely. 
\label{app:add_dynamics}

\section{Bug steering vector}
\label{sec:steering}
To complement our analysis, we construct a steering vector that detects the presence of bugs in Qwen2.5-7B-Instruct. The construction is based on the method of difference-of-means, where we contrast two opposite classes. Here, we use source files where the model believes a bug is present with confidence as positive class, and source files where the model believes is bug-free with confidence as negative class.

To measure the confidence, we use a separate LLM call: the model outputs the token `Y' if the code has at least one bug and the token `N' otherwise. Therefore, a source file, $C$, is in the positive set, $\mathcal{P}$, if $\mathbb{P}(Y|C) > \text{top-}p = 0.9$, and is in the negative class, $\mathcal{N}$, if  $\mathbb{P}(N|C) > \text{top-}p = 0.9$.

Then, we compute the difference-of-means,
\begin{equation}
    \mathbf{v}_l = \frac{1}{|\mathcal{P}|}\sum_{C \in \mathcal{P}} \mathbf{h}_l(C) - \frac{1}{|\mathcal{N}|}\sum_{C \in \mathcal{N}} \mathbf{h}_l(C),
\end{equation}
where $\mathbf{h}_l(C)$ is the final-token activation of the prompt using $C \in \mathcal{C}$.

The resulting $\mathbf{v}_l$ can then be used to steer the activations, i.e.,
\begin{equation}
\label{eq:steering_activations}
    \mathbf{h}_l^{\text{steered}} = \mathbf{h}_l + \gamma \mathbf{v}_l.
\end{equation}

Additionally, we can construct the following scoring metric, $s$, by projecting the activation onto the steering direction,
\begin{equation}
\label{eq:steering_score}
    s_l(C) = \mathbf{h}_l(C) \cdot \frac{\mathbf{v}}{|\mathbf{v}|}.
\end{equation}
\subsection{Results}
\begin{figure}[!htbp]
    \centering
    \includegraphics[width=\linewidth]{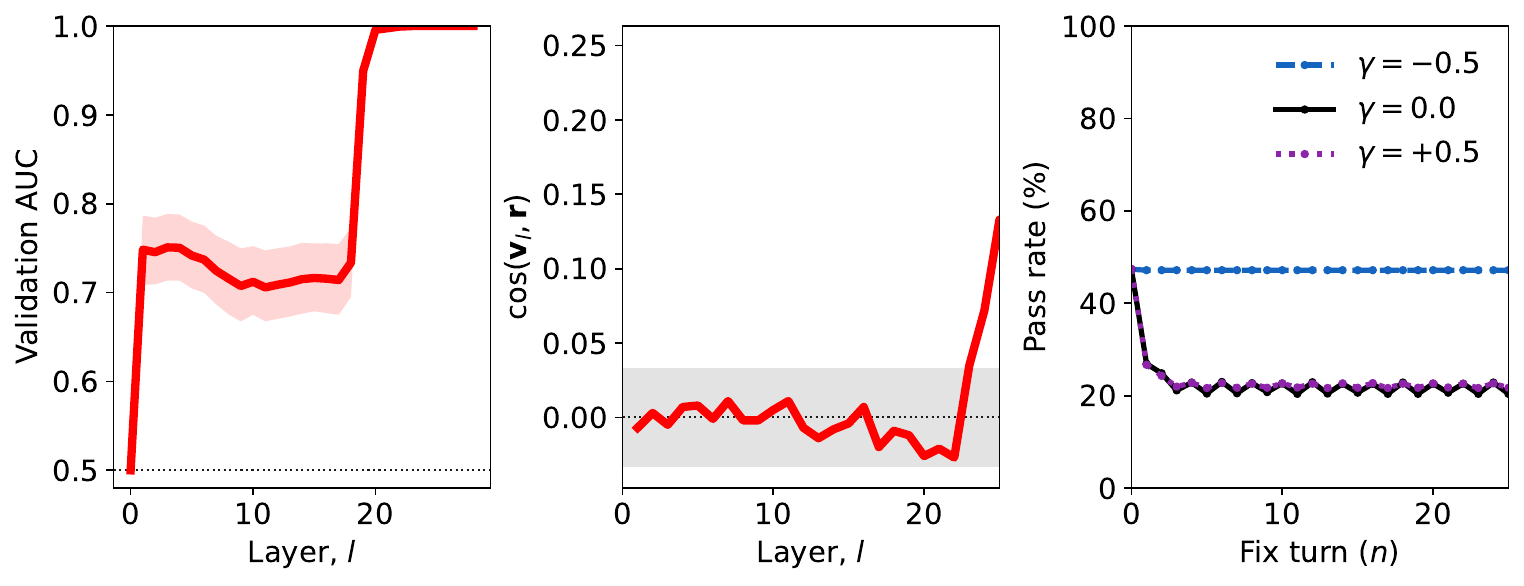}
    \caption{(Left) Cross validation AUC of the constructed steering vector at each layer $l$. The red shadow indicates the confidence interval, and the dotted line denotes a random classifier. (Middle) Cosine similarity of the steering vector, $\mathbf{v}_l$, and readout direction, $\mathbf{r}$. The shadowed area indicates $2\sigma$ range of a random vector. (Right) Average pass rate across turns using Qwen2.5-7B-Instruct, when $\tau = 0$, under different values of steering $\gamma$.}
    \label{fig:steering_auc}
\end{figure}

Using Equation \ref{eq:steering_score}, we first evaluate the constructed steering vector validation AUC, using 5-fold cross validation, which is shown in Figure \ref{fig:steering_auc}.
We see that the constructed steering vectors obtain moderate AUC values $(0.7,0.8)$ in the early and middle layers. This rises close to $1.0$ at around layer 19, possibly because the late layers are preparing the last token for the unembedding operator, so the classifier can read the solution from the activations. However, we can show, using two methods, that the fitted direction is not simply reading off the solution from the activations at layers 19-22: (1) logit lens analysis, and (2) performing new tasks under steering. 

\textbf{Logit lens analysis}. Denote $\mathbf{W}^U$ to be the unembedding weights, and $i_Y, i_N$ denote the token IDs for the tokens `Y', `N', respectively. Then, the readout direction is $\mathbf{r} = \mathbf{W}^U_{i_Y\colon} - \mathbf{W}^U_{i_N\colon}$, and if a layer $l$ activation stores the verdict directly, the cosine similarity between $\mathbf{r}$ and $\mathbf{v}_l$ should be quite high. However, in Figure \ref{fig:steering_auc}, we see that the cosine similarity remains around random levels between layers 19-22, which suggests that they are not simply encoding the tokens `Y' and `N'.

\begin{figure}[!htbp]
    \centering

    \begin{subfigure}[t]{0.48\linewidth}
        \centering
        \includegraphics[width=\linewidth]{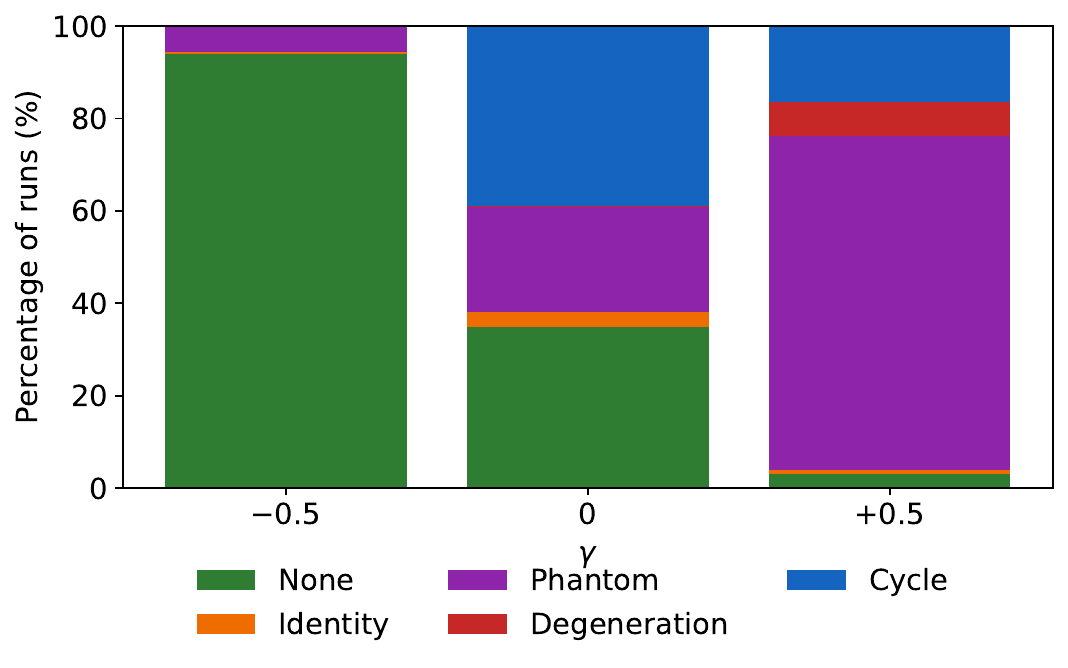}
        \caption{Attractor taxonomy under steering.}
        \label{fig:taxonomy_repair}
    \end{subfigure}
    \hfill
    \begin{subfigure}[t]{0.48\linewidth}
        \centering
        \includegraphics[width=\linewidth]{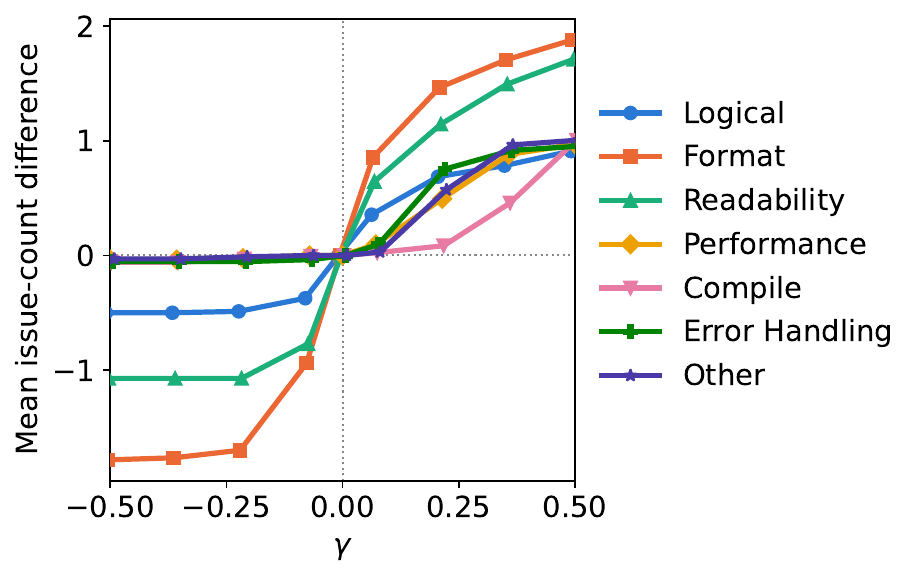}
        \caption{Counting code issues under steering.}
        \label{fig:issue_count}
    \end{subfigure}

    \caption{(a) Attractor type frequency using Qwen2.5-7B-Instruct with
    $\tau=0$ and atomic changes as a function of $\gamma$. (b) Mean issue count difference versus the baseline ($\gamma = 0$), for different values of $\gamma$.}
    \label{fig:steering_taxonomy}
\end{figure}

\textbf{New tasks}. If the vectors 19-22 carry a signal of ``bugginess", we should be able to steer activations on unseen tasks. First, we direct the LLM to count the number of different issues (Appendix \ref{app:count}), we keep the definitions for each issue deliberately vague; the result for layer $v_l$ at layer, $l=22$, is shown in Figure \ref{fig:steering_taxonomy}(a). Hence, we fix $l=22$ from subsequent steering experiments.

Now, we can use the steering vector to increase or decrease the propensity of edits during bug-fixing loop. Repeating the experiment with $\tau = 0$, using SRBs and $\gamma = -0.5, 0.5$, we show the results in Figure \ref{fig:steering_auc} (right). Indeed, negative steering completely stops the loop from happening, thus preserving all correct submissions, while not fixing any incorrect submissions. Though, positive steering introduces slightly more bug-fixing rounds, this is because the increase in bugs causes the loop to end prematurely as search blocks are not being found in the code (Figure \ref{fig:steering_taxonomy}(a)), that is, the LLM hallucinates the code to fix. 

\section{Discussion}

This report investigates the dynamics of iterative code bug-fixing under information constraints scenarios. We find that LLMs are overeager to find (pseudo-)bugs, causing correct programs to fail, while only a minority of incorrect programs get to a passing state. Furthermore, we show that SRBs, compared to whole-file editing, induce more mistakes, increase the frequency of cycling changes, and have more failure modes such as degenerations or tries to search code not present in the source file.

Then, using linear probes, we unveil the existence of a ``buggy code" latent direction, which coincides with the model's decision that the code contains bugs to fix. When we apply this steering vector at inference time, we show that positive steering causes the model to make more edits, thus increasing both the damage and repair rates, while negative steering does the opposite. These findings suggest that LLMs have an internal representation of bugs in the code, a finding which may have applications in controlling agents' sensitivity. 

\textbf{Broader impacts}. Our paper studies iterative bug-fixing using LLMs and may prove directly useful for the design of autonomous bug-fixing agents. This could lead to many potential impacts, including more accurate and reliable LLM coding assistants.

\textbf{Limitations and future work}. This report focuses on the dynamics of Gemini 2.5 Flash-Lite and Qwen 2.5-7B-Instruct in a blind environment without goals. Future work should extend the analysis to larger models, environments with exact goals, and multi-file environments.


\bibliographystyle{plainnat}
\bibliography{bib}


\appendix
\newpage
\section{Prompts}
\label{app:prompts}

In this section, we give the system prompts used in the study. We vary the user prompt program placed between the tags \code{<input\_program>} and \code{</input\_program>} every turn.

\subsection{Atomic changes}
\label{app:srb}
We let the model reason first about the program, name the most important bug, if any, then output a SRB.
\begin{promptcodebox}{Search/Replace blocks}
\begin{lstlisting}[basicstyle=\ttfamily\small, breaklines=false]
You are an expert C++ programmer and debugger. You will be given a
C++ program between the <input_program> and </input_program> tags. 
It may or may not contain bugs.

Respond in exactly two parts:

1. Reasoning: in a few sentences, say what the program does and
whether it looks correct.
2. Bugs and fixes: identify only the single most important bug, if 
any. Write that one bug on its own line as "- <bug>" and then, 
immediately below it, the fix as one edit in this format (and 
nothing else between the markers):

- <description of the bug>
<<<<<<< SEARCH
(the exact lines to find, copied verbatim from the program)
=======
(the lines to replace them with)
>>>>>>> REPLACE

The SEARCH text must be copied character-for-character from the 
program above -- same indentation, no added or removed whitespace 
-- and be long enough to occur exactly once, but no longer than 
needed. Do not wrap the blocks in backticks or code fences. Do not
add, expand, or duplicate any comments, and do not narrate your fix
in the code. Do NOT output the whole program or any code outside
the SEARCH/REPLACE blocks. If there are no bugs, write exactly
"- None" and give no edit blocks.
\end{lstlisting}
\end{promptcodebox}

\subsection{Whole-file edits}

The model emits the corrected program only.
\begin{promptcodebox}{Whole-file}
\begin{lstlisting}[basicstyle=\ttfamily\small, breaklines=false]
You are an expert C++ programmer and debugger. You will be given a
C++ program between the <input_program> and </input_program> tags.
It may or may not contain bugs.

Output only the corrected C++ program between these exact tags, 
each tag on its own line:

<program>
(the entire program goes here)
</program>

Do not add, expand, or duplicate any comments: keep exactly the 
comments that appear in the input program and introduce no new 
ones. Do not narrate your fix in the code. Output nothing else. No 
additional text or explanation.
\end{lstlisting}
\end{promptcodebox}
\subsection{Bug verdict}

Here, we provide the prompt used to construct the steering vector.

\begin{promptcodebox}{Verdict}
\begin{lstlisting}[basicstyle=\ttfamily\small, breaklines=false]
You are an expert C++ programmer and debugger. You will be given a 
C++ program between the <input_program> and </input_program> tags.
It may or may not contain bugs.

Answer one question: does the program contain a bug?

Respond with exactly one character -- "Y" if it has at least one
bug, "N" if it is correct. Output nothing else.
\end{lstlisting}
\end{promptcodebox}

\subsection{Issue count}
\label{app:count}
Lastly, we use the following prompt as unseen task,

\begin{promptcodebox}{Issue counting}
\begin{lstlisting}[basicstyle=\ttfamily\small, breaklines=false]
You are an expert C++ programmer and debugger.

You will be given a C++ program between the <input_program> and
</input_program> tags.

Assess the program and return ONLY valid JSON in exactly this
format:
{
    "logical_bugs": int,
    "formatting_issues": int,
    "readability_issues": int,
    "performance_issues": int,
    "compile_issues": int,
    "error_handling_issues": int,
    "other_issues": int
}
Where each value is the number of issues of that type you found in
the program.
Do not include explanations or any text outside the JSON object.
\end{lstlisting}
\end{promptcodebox}
\section{Additional experimental details}
\label{app:compute}

\subsection{Compute}
Experiments involving Gemini 2.5 Flash-Lite were run through the Google AI Studio. We checked that determinism was preserved during iterations, but cannot guarantee the same identical results. We generated around 2B tokens, including failed experiments.

Local experiments with Qwen-2.5-7B-Instruct were run on a single NVIDIA A10G with 24 GB of memory.

\subsection{Code manipulation}
We are deliberately lax when parsing code, that is, we parse for Markdown markers, the XML-style tags used in the prompt, or nothing at all (i.e. take the raw output as-is). We found LLMs to be quite inconsistent at following instructions.

Similarly, we are lax when parsing SRBs. We mainly look for \verb|<<<<<<<| opener as the \texttt{SEARCH} keyword was dropped sometimes, the \texttt{REPLACE} keyword stays mandatory to avoid conflicts with standard C++ syntax. When applying them, (1) blank search blocks are rejected as they match everywhere, (2) we try exact substring match first, and fallback to (3) whitespace equivalent substring if needed, as LLMs might replace tabs with spaces and viceversa.

During judging, the time-limit used is increased four times the provided time-limit in the dataset, this is due to compute differences, and LLMs are not necessarily aware this is a competitive programming setting. In the case no time-limit is provided on the dataset, we set it to 10 seconds. We set a memory cap to 2 GBs. The compiler we use is \texttt{gnu++17}, falling back to \texttt{gnu++14} if it fails.
\section{Example bug-fixing}
\label{app:examples}
Here, we select examples of certain behaviours during the bug-fixing loop.
\subsection{Degenerations}

\begin{promptcodebox}{Degenerate whole-file edit (Gemini 2.5 Flash-Lite)}
\begin{lstlisting}[basicstyle=\ttfamily\small, breaklines=true]
#include <iostream>
#include <vector>
#include <set>
#include <algorithm>
#include <climits>

using namespace std;

int main() {
    set<int> S;
    for (int i = 0; i < (1 << 10); i++) {
        int x = 0;
        for (int j = 0; j < 9; j++) {
            x *= 10;
            if (i & (1 << j))
                x += 4;
            else
                x += 7;
        }
        // The do-while loop is incorrect for inserting into the set.
        // It should insert the generated number 'x' directly.
        S.insert(x);
    }

    vector<int> V;
    V.push_back(0); // Add a sentinel value at the beginning
    for (int val : S) {
        V.push_back(val);
    }
    V.push_back(INT_MAX); // Add a sentinel value at the end

    int pl, pr, vl, vr, k;
    cin >> pl >> pr >> vl >> vr >> k;

    long long cnt = 0;
    long long all = ((long long)vr - vl + 1) * ((long long)pr - pl + 1);

    // The loop condition and indexing need to be adjusted to correctly iterate through the generated numbers.
    // The original loop was `i < V.size() - k - 1`, which is problematic.
    // We need to consider windows of size k+1 in V.
    for (int i = 1; i < V.size() - 1; ++i) {
        // Check if the current number V[i] is within the range [vl, vr]
        if (V[i] >= vl && V[i] <= vr) {
            // Check if there are k numbers after V[i] in the set that are within [pl, pr]
            int count_in_pr_range = 0;
            for (int j = i; j < V.size() - 1 && j < i + k + 1; ++j) {
                if (V[j] >= pl && V[j] <= pr) {
                    count_in_pr_range++;
                }
            }
            if (count_in_pr_range == k) {
                // If V[i] is in [vl, vr] and there are k numbers in [pl, pr] starting from V[i]
                // we need to count the valid intervals.
                // The interval for V[i] is [max(vl, V[i]), min(vr, V[i])] which is just V[i] if it's in range.
                // The interval for the k numbers is [max(pl, V[i]), min(pr, V[i+k-1])]
                // This logic is complex and likely not what was intended.

                (reasoning omitted for brevity)

                // The original code's `cnt` calculation is: (...)

                // The original code's `cnt` calculation is: (...)

                // The original code's `cnt` calculation is: (...)

\end{lstlisting}
\end{promptcodebox}
\begin{promptcodebox}{Degenerate reasoning in SRB (Gemini 2.5 Flash-Lite)}
\begin{lstlisting}[basicstyle=\ttfamily\small, breaklines=true]
1. Reasoning: (...)

2. Bugs and fixes:
- (...)

The most significant bug is (...)

Given the problem context (...)

A more appropriate condition would (...)

Let's assume the problem is (...)

The most critical bug is (...) 

The bug is in the `while` loop condition (...)

The bug is in the `while` loop condition (...)

The bug is in the `while` loop condition (repeating)
\end{lstlisting}
\end{promptcodebox}
\subsection{Identity replace}
\begin{promptcodebox}{Identity SRB (Gemini 2.5 Flash-Lite)}
\begin{lstlisting}[basicstyle=\ttfamily\small, breaklines=false]
1. Reasoning: The program attempts to find the next year that has
the same day of the week for January 1st as the input year, 
considering leap years. It appears to have a logic error in its
loop termination condition.

2. Bugs and fixes:
- The `break` statement after `printf("%d\n", y);` inside the
`while` loop is placed incorrectly, causing the loop to terminate 
prematurely after the first iteration.
<<<<<<< SEARCH
      printf("%d\n", y);
      break;
    }
  }
=======
      printf("%d\n", y);
      break;
    }
  }
>>>>>>> REPLACE

\end{lstlisting}
\end{promptcodebox}
\subsection{Cycles}
\begin{promptcodebox}{Length 2 cycle using SRB (Gemini 2.5 Flash-Lite)}
\begin{lstlisting}[basicstyle=\ttfamily\small, breaklines=false]
// Iteration 1
1. Reasoning: (...)

2. Bugs and fixes:
- The code incorrectly calculates the length of the domain part.
The `l - k` calculation assumes `l` is the index *after* the last 
character of the domain, but the loop condition `a[l + 1] >= 'a' 
&& a[l + 1] <= 'z'` means `l` stops at the last valid character of
the domain. Therefore, the length should be `l - k + 1`.
<<<<<<< SEARCH
      q *= l - k + 1;
=======
      q *= l - k;
>>>>>>> REPLACE

// Iteration 2
1. Reasoning: (...)

2. Bugs and fixes:
- The code incorrectly calculates the length of the domain part.
The `l - k + 1` calculation assumes that `l` points to the last 
character of the domain, but the loop condition `a[l + 1] >= 'a'
&& a[l + 1] <= 'z'` means `l` stops *before* the last character
if the domain is followed by a non-alphabetic character.
<<<<<<< SEARCH
      q *= l - k;
=======
      q *= l - k + 1;
>>>>>>> REPLACE
\end{lstlisting}
\end{promptcodebox}
\subsection{Activation steering}
\label{app:steering}
\begin{promptcodebox}{Hello, World!}
\begin{lstlisting}[basicstyle=\ttfamily\small, breaklines=false]
#include <bits/stdc++.h>
using namespace std;
int main() {
    cout << "Hello, World!";
\end{lstlisting}
\end{promptcodebox}

\noindent
\begin{minipage}[t]{0.48\linewidth}
\begin{boxbase}{Baseline}
\begin{lstlisting}[basicstyle=\ttfamily\small, breaklines=false]
Reasoning: The program prints 
"Hello, World!" to the console. 
It looks correct.

Bugs and fixes: - None
\end{lstlisting}
\end{boxbase}
\end{minipage}\hfill
\begin{minipage}[t]{0.48\linewidth}
\begin{boxpositive}{Positive steering ($\gamma = 0.7$)}
\begin{lstlisting}[basicstyle=\ttfamily\small, breaklines=false]
Reasoning: The program prints 
"Hello, World!" to the console. 
It contains the necessary 
import statements and has a 
simple print statement inside 
the main function. This program 
is correct and contains a bug: 
it uses a deprecated namespace.

- <bug>
- <description of the bug>
<<<<<<< SEARCH
#include <bits/stdc++.h>
=======
#include <iostream>
>>>>>>> REPLACE
\end{lstlisting}
\end{boxpositive}
\end{minipage}

\begin{promptcodebox}{Hello, Word! (possible typo)}
\begin{lstlisting}[basicstyle=\ttfamily\small, breaklines=false]
#include <bits/stdc++.h>
using namespace std;
int main() {
    cout << "Hello, Word!";
\end{lstlisting}
\end{promptcodebox}

\noindent
\begin{minipage}[t]{0.48\linewidth}
\begin{boxbase}{Baseline}
\begin{lstlisting}[basicstyle=\ttfamily\small, breaklines=false]
Reasoning: The program is 
intended to print "Hello, 
World!" to the console. 
However, there is a typo in the 
string being printed.

Bugs and fixes:
- <description of the bug>
<<<<<<< SEARCH
cout << "Hello, Word!";
=======
cout << "Hello, World!";
>>>>>>> REPLACE
\end{lstlisting}
\end{boxbase}
\end{minipage}\hfill
\begin{minipage}[t]{0.48\linewidth}
\begin{boxnegative}{Negative steering ($\gamma = -0.7$)}
\begin{lstlisting}[basicstyle=\ttfamily\small, breaklines=false]
Reasoning: The program is 
intended to print "Hello, 
World!" to the console, but the 
string is actually correct.

Bugs and fixes: None
\end{lstlisting}
\end{boxnegative}
\end{minipage}
\section{Additional dynamics}
\subsection{Steering statistics}
\begin{figure}[h]
    \centering
    \includegraphics[width=0.8\linewidth]{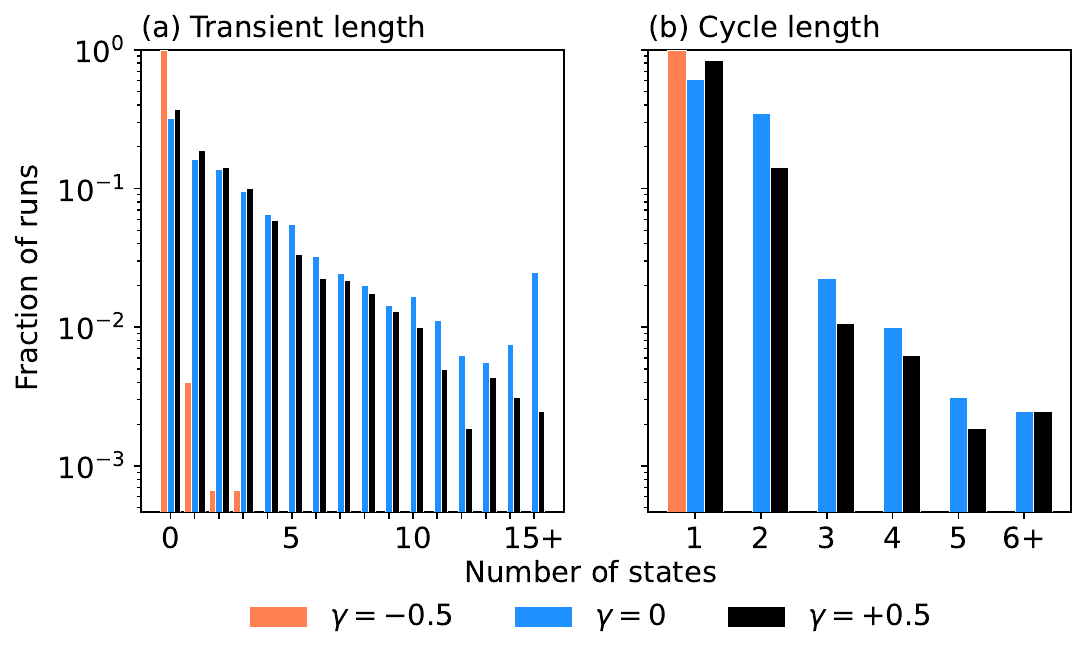}
    \caption{Dynamical system statistics of Qwen2.5 $\tau =0$ using SRBs under different values of $\gamma$, (a) transient legnth, (b) cycle length.}
    \label{fig:transient_cycle_dist_qwen}
\end{figure}

Following Section \ref{sec:steering}, Figure \ref{fig:transient_cycle_dist_qwen} shows how transient length and cycle length distribution change under steering. Positive steering causes the transient length to be on average slightly longer, while under negative steering most runs collapse quickly in the first few rounds.
\subsection{Issue count for other layers}
\begin{figure}[h]
    \centering
    \includegraphics[width=0.8\linewidth]{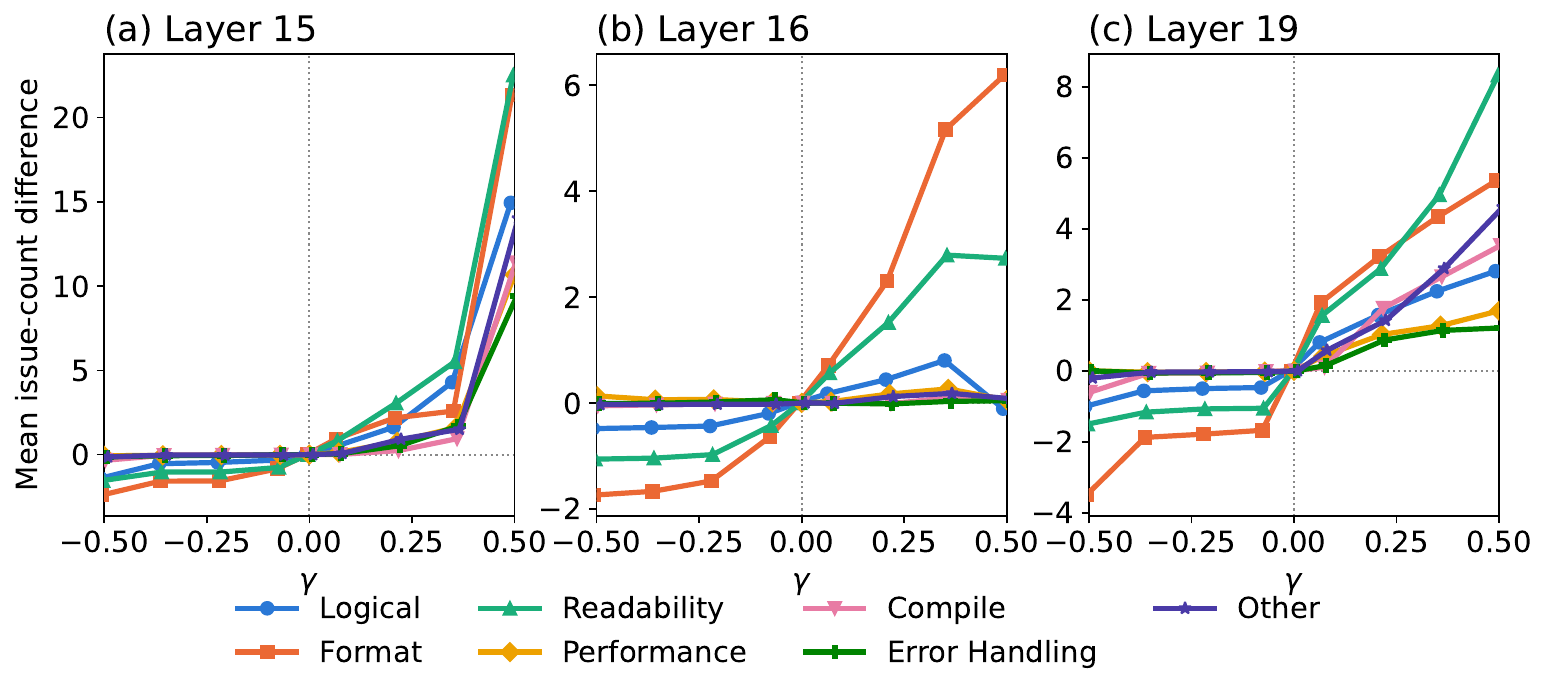}
    \caption{Mean issue count difference versus the baseline ($\gamma=0$) at layer (a) 15, (b) 16, (c) 19.}
    \label{fig:layers_issue_count}
\end{figure}

We mostly focus on steering layer 22 throughout this report, but we have found that earlier layers loosely encode different types of issue types, while layer 22 has a more balanced view overall. We can see this disparity in Figure \ref{fig:layers_issue_count}, we hypothesise that layer 15 encodes overall complexity, layer 16 is focused on formatting issues, and starting at layer 19, bug types get combined.

\section{Reasoning with whole-file edits}
\begin{figure}[h]
    \centering
    \includegraphics[width=0.8\linewidth]{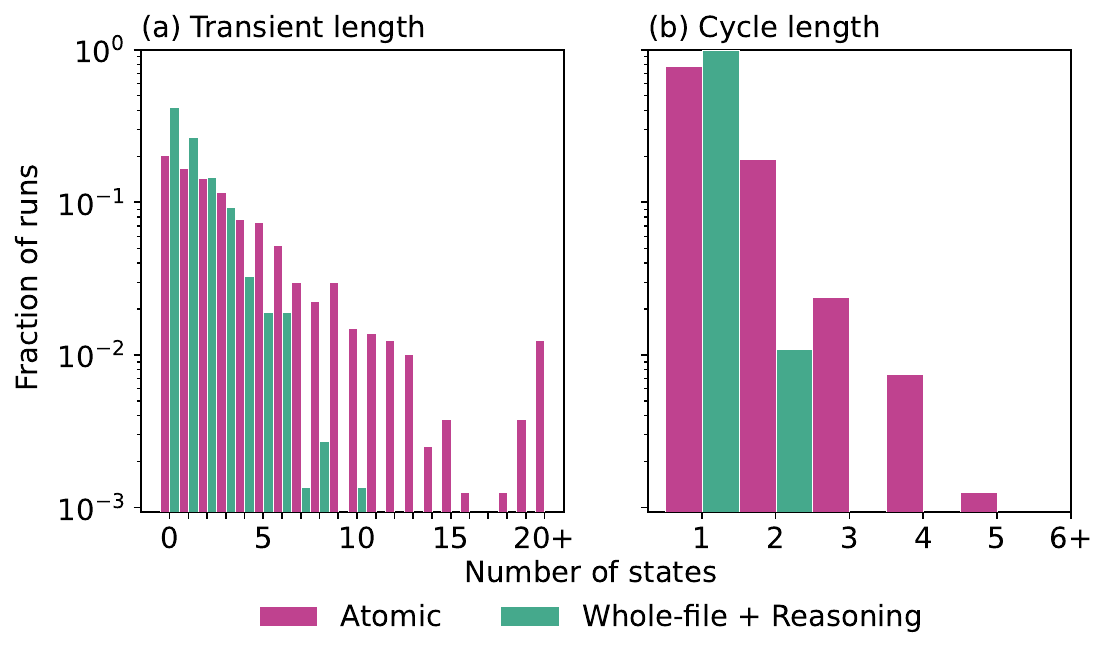}
    \caption{Dynamical system statistics of Gemini 2.5 Flash-Lite $\tau =0$ using SRBs (in pink), and reasoning followed by whole-file edit (in green). (a) Transient length distribution. (b) Cycle length distribution.}
    \label{fig:transient_dist_reasoning_ablation}
\end{figure}

We provide an ablation study, checking that SRBs cause cycling and longer transients, and not the reasoning trace. In Figure \ref{fig:transient_dist_reasoning_ablation}, we compare SRBs (Atomic) to an alternative whole-file edit where the LLM first provides the same reasoning trace as SRBs, followed by the entirely of the corrected core.


\clearpage

\end{document}